\documentclass[11pt,a4paper,english]{article}
\usepackage[utf8]{inputenc}
\usepackage[T1]{fontenc}
\usepackage{cmap} 
\usepackage{babel}
\usepackage{graphicx}
\usepackage{amssymb}
\usepackage[table]{xcolor}
\usepackage[hyphens]{url}
\usepackage[hidelinks]{hyperref}
\hypersetup{breaklinks=true}
\usepackage[authoryear, round]{natbib}
\usepackage{longtable}
\usepackage{tabularx}
\usepackage{hhline}
\usepackage{caption} 
\usepackage{authblk}
\date{\today}
\title{Roadless space is greatly diminished by logging in intact forest landscapes of the Congo Basin}

\author[1,2]{Fritz Kleinschroth\thanks{To whom correspondence should be addressed. Email: fritz.kln@gmail.com}}
\author[2]{John R. Healey}
\author[1]{Sylvie Gourlet-Fleury}
\author[1]{Frederic Mortier}
\author[3,4]{Radu S. Stoica}
\affil[1]{CIRAD, Biens et Services des Ecosystems Forestiers Tropicaux, 34398 Montpellier, France}
\affil[2]{Bangor University, School of Environment, Natural Resources and Geography, Bangor Gwynedd, LL572UW, UK}
\affil[3]{Universite de Lille 1, Laboratoire Paul Painleve, 59655 Villeneuve d'Ascq, France; }
\affil[4]{Institut de Mecanique Celeste et Calcul d'Ephemerides, Observatoire de Paris,	75014 Paris, France}

\begin{document}
	\maketitle

	\begin{abstract}
		Forest degradation in the tropics is often associated with roads built for selective logging. The protection of Intact Forest Landscapes (IFL) that are not accessible by roads is high on the biodiversity conservation agenda, a challenge for logging concessions certified by the Forest Stewardship Council (FSC). A frequently advocated conservation objective is to maximise the retention of “roadless space”, a concept that is based on distance to the nearest road from any point. We developed a novel use of the empty space function -- a general statistical tool based on stochastic geometry and random sets theory -- to calculate roadless space in a part of the Congo Basin where there has recently been rapid expansion of road networks. We compared the temporal development of roadless space in certified and non-certified logging concessions inside and outside areas declared as IFL in the year 2000. Between 1999 and 2007, rapid road network expansion led to a marked loss of roadless space in IFL. After 2007, this trajectory levelled out in most areas, due to an equilibrium between newly built roads and abandoned roads that became revegetated. However, concessions within IFL that have been certified by FSC since around 2007 showed continued decreases in roadless space, thus reaching a level comparable to all other concessions. Only national parks remained road-free. We recommend that forest management policies make the preservation of large connected forest areas a top priority by effectively monitoring -- and limiting -- the occupation of space by roads that are permanently accessible.
	\end{abstract}

\section{Introduction}
	Road networks are expanding rapidly around the world, connecting people and resources, increasingly in remote regions \citep{Laurance2014c}. This provides a huge challenge for species conservation as roads can act as a physical barrier to migration and therefore potentially limit gene flow, reducing effective population sizes \citep{Benitez-Lopez2010,Laurance2015}. Furthermore, roads can be corridors for species invasions into remote landscapes, with people and their vehicles acting as dispersal vectors \citep{VonDerLippe2007, Veldman2010}. Consequently, roadlessness increases overall landscape connectivity for most forest species \citep{Crist2005} and has been successfully used as a measure, e.g., to predict species richness and composition of Amazonian bird communities \citep{Ahmed2014}. Inventoried roadless areas (IRA) became part of forest and conservation legislation in 1999 in the USA due to their effectiveness for conservation outside protected areas \citep{DeVelice2001}. Laurance et al. \citep{Laurance2014c} extended this approach by proposing a global strategy to regulate road building.
	
	In tropical regions forest degradation, unregulated hunting, and deforestation due to agricultural colonization have been associated with roads built for selective logging \citep{Wilkie2000,Brandt2016}. Therefore, forest areas that are not accessible by roads are considered of highest conservation value because they provide habitats that are not immediately impacted by major human activities. The protection of such "Intact Forest Landscapes" (IFL) that are not penetrated by roads is high on the biodiversity conservation agenda. They have been defined for the year 2000 as those areas $>$~500~km\textsuperscript{2} and $>$~10~km wide that are outside a buffer of one~km around any road or settlement \citep{Potapov2008}. While ecologically the intactness of a forest depends on many factors, in remote tropical regions the operational use of the term "intact" corresponds to the concept of roadlessness. The underlying assumption is that important impact of roads inside intact forest landscapes are not only the dissection of formerly connected habitats but also the process of incision \citep{Jaeger2000} that opens the forest for anthropogenic disturbances \citep{Laurance2009a}. Due to the easy detectability of newly constructed roads in otherwise closed canopy forests, the current identification of intact forests excludes any forest that has recently been penetrated by roads built for selective logging, independent of harvest intensity. However, e.g. in Central Africa, only 20\% of roads built for logging are permanently accessible, with the remaining 80\% becoming rapidly vegetated \citep{Kleinschroth2015}.
	
	Forest certification, such as that of the Forest Stewardship Council (FSC), provides market-based incentives for logging operators who are audited for their sustainable forest management implying reduced impact logging standards, traceability of timber and social welfare for workers \citep{Blackman2011}. Adherence to these standards should require the prevention of long-term negative impacts on forest ecosystems, e.g. through poorly designed roading systems \citep{FSC2010}. Under increasing pressure from the international environmental NGO Greenpeace, FSC has recently passed a motion to better protect IFL as part of their "high conservation value forest" policy and implement this as a new standard by the end of 2016 \citep{Rodrigues2014}. Central Africa, with some of the world's least exploited tropical forests, is at the center of attention for this policy change. 
	
	To quantify roadlessness in forest landscapes, new methods are required. This is essential to determine if logging operations certified for their good management did limit road network expansion. In the Sangha river catchment, a prime target area for selective logging activities in the Congo Basin, we assessed the change in logging road networks between 1999 and 2015 in order to determine how this differed between forest areas varying in certification status and location inside or outside IFL. In order to quantify and test the spatial distribution of road networks, we developed a novel use of the empty space function (\textit{F}), an established mathematical tool that allows quantification and testing of the spatial distribution of roadless areas. We hypothesized that roadless space would have increased less rapidly in FSC-certified than in non-certified concessions, and even less inside intact forest landscapes, assuming a positive interaction between certification and IFL. Based on our findings we discuss the implications for forest conservation and management.

	\section{Methods}
	\subsection{Study area and data collection}
\begin{figure}[!htbp]
			\centering
			\includegraphics[width=1\textwidth]{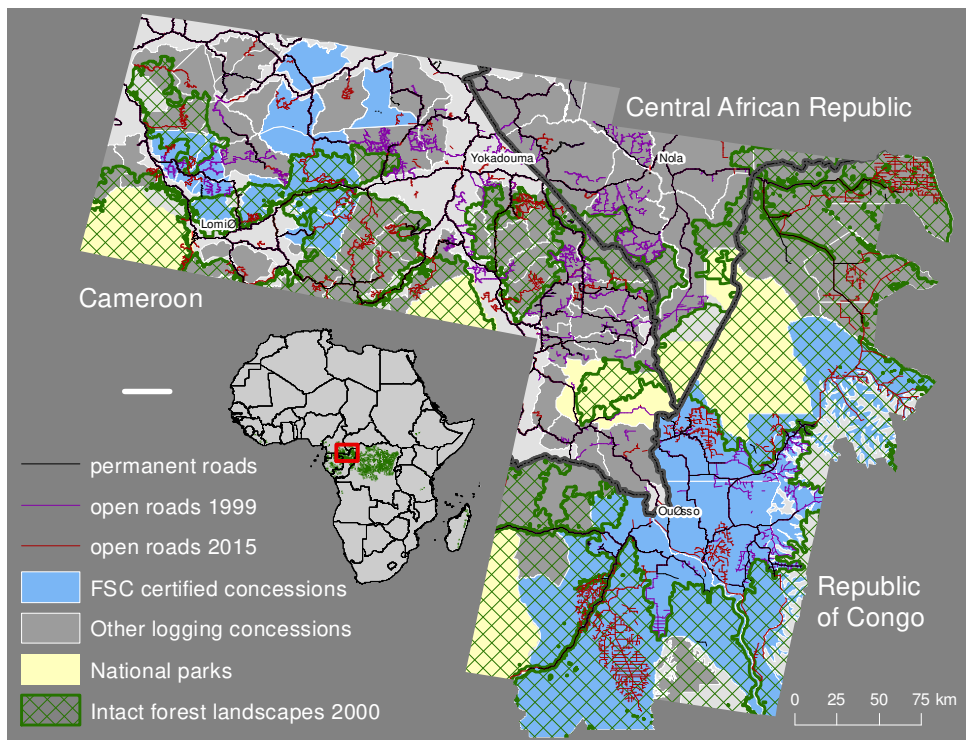}
			\caption{\label{fig:map} Overview of the study area and its location on the African continent (inset). National Parks, Forest Stewardship Council certified and non-certified concessions, and their spatial overlap with intact forest landscapes as defined by Potapov et al. \citep{Potapov2008} for the year 2000 are shown overlain by permanent roads and those that were open in 1999 and in 2015.} 
		\end{figure}

	The study area is 107~000~km\textsuperscript{2} in extent, covering parts of Republic of Congo, Cameroon and Central African Republic. Most of the area is characterized by Guineo-Congolian semi-deciduous forest \citep{White1983}. This region has only recently been subject to a rapid expansion of logging road networks into previously intact forests \citep{Laporte2007b}. These logging operations mostly take place in concessions, large state-owned forest areas that are allocated to companies for timber harvesting according to a forest management plan \citep{Lescuyer2015}. We analysed a total of 67 concessions of varying sizes (range 217-10~270, median 660~km\textsuperscript{2}). These are operated by 17 different companies, mostly in groups of several bordering concessions (range 365-15~560, median 1970~km\textsuperscript{2}) to reduce infrastructure costs. Between 2006 and 2009 14 concessions have been awarded FSC certification \citep{Bayol2012}, covering 40\% of the total area of concessions. These certified concessions are operated by five different companies of which three have almost all of their concession area certified and two have only one third certified. Fifty-five percent of the study area was classified as intact forest landscapes for the year 2000 \citep{Potapov2008}. We classified the overall area into five different management categories: logging concessions certified or not certified by FSC, each inside and outside IFL, and national parks (Fig.~\ref{fig:map}).
	
\begin{figure}[!htbp]
			\centering
			\includegraphics[width=1\textwidth]{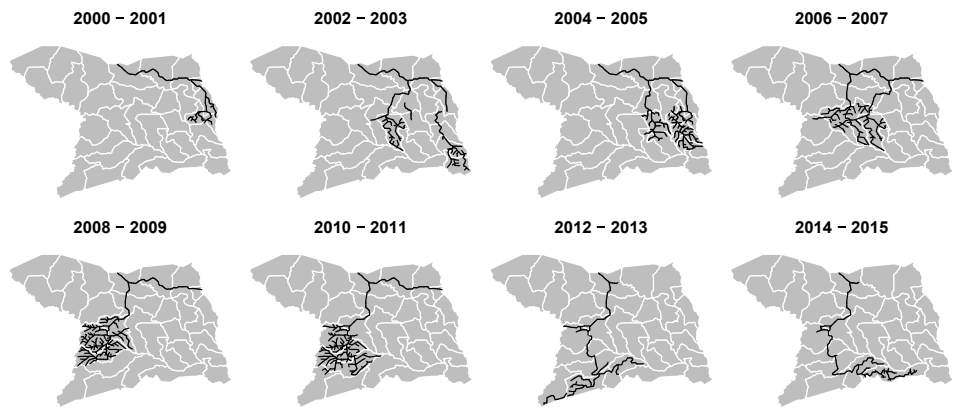}
			\caption{\label{fig:aac} Temporal development of the road network in one logging concession in Cameroon. White lines delimit the division of the concession into 30 annual felling areas (AAC's), black lines show roads that were open during the respective two-year intervals. The east-west extension of the concession is ca. 50~km}. 
		\end{figure}

	Given the economic dominance of logging in the study area, we associate the majority of roads constructed with timber extraction. Logging road networks are highly dynamic, with 50\% of roads persisting on Landsat imagery for less than four years due to vegetation recovery \citep{Kleinschroth2015}. We vectorised forest roads during nine two-year time intervals based on a time series of 222 LANDSAT~7 and 8 images captured between 1999 and 2015. Based on the contrasting spectral properties of bare soil and recovering vegetation we were able to differentiate open (actively used) and abandoned (in process of revegetation) roads \citep{Kleinschroth2015}. In the present study we included only those roads that were open on any image during the 16-year time interval. The majority of logging concessions in the region are operated under a management plan, effectively limiting the amount of wood harvested annually by demarcating annual felling areas (assiettes annuelles de coupe, AAC) based on timber inventories \citep{Karsenty2008}. These AAC's are the most important factor determining where roads are built each year (Fig.~\ref{fig:aac}).
	
	To verify the accuracy of our methodology we first tested it on river networks. These were derived from a digital elevation model \citep{Lehner2006} and thus directly reflect the topography. Despite the fundamental difference in ecological characteristics between rivers and roads, both are general random sets and can be analysed with the same tool. A priori the distribution of the rivers in the region should be independent of the management and intactness of forests. Therefore \textit{F} is expected to show no difference for rivers between zones of different forest management type (because rivers are not managed in that region) but to vary for roads due to the effects of management. We duly found that empty space curves for river networks were very similar throughout all forest management categories (Fig.~\ref{fig:riv}). Pairwise \textit{t} tests showed no significant differences: \textit{P} values were $>$~0.9 for all combinations. This does not mean that the river networks were identical but it is consistent with the assumption that the distribution of rivers is independent of forest management across the study area. This indicates that the methodology is accurate in that it does not show variation between equally distributed patterns such as the existing river network.
	
	\subsection{Quantifying roadlessness with the Empty-Space Function}
	The method most commonly used to evaluate intactness of forests is based on one pre-defined buffer distance around any road or settlement \citep{Potapov2008,Tyukavina2015,Herold2011}. This approach is quick and easy but lacks accuracy in that it does not take into account the highly dynamic nature of forest degradation \citep{Goetz2015}. Different species have different radii of movement just as different human land uses affect forest functions over varying distances \citep{ Coffin2007396}. The binary classification of an area based on a buffer (intact vs. degraded) is closely linked to road length density and does not take road location in the overall landscape into account. An alternative way to characterize landscapes is to quantify, for a certain area, the distance from each point to the nearest road \citep{Riitters2003}. This idea is implemented in the metric of roadless volume \citep{Watts2007} where a pixel of an area is assumed to have a higher value the further away it is from a road, which then allows calculation of the volume under this pseudo-topographic surface as an index of roadlessness. We follow the approach of Riitters \& Wickham \citep{Riitters2003} by integrating their idea in an established mathematical function.
	
	The Empty-Space function $\mathit{F}$ (or spherical contact distribution function) is based on stochastic geometry and random sets theory \citep{Lieshout1996,Foxall2002,Gelfand2010}. The main hypothesis for our work is that that the road networks is the realization of a random set. For a stationary random set $\mathit{X}$, the distance from an arbitrary point $u \in \mathbb{R}^2$ to the nearest element of $\mathit{X}$ is $dist(u,\mathit{X})$. For a given radius $\mathit{r}$, $\mathit{F}$ can be described as $\mathit{F}(r)=\mathcal{P}(dist(u,\mathit{X})\leq r)$. Given the assumption of stationarity this does not depend on $\mathit{u}$. Similar to random variables, the $\mathit{F}$ function can be interpreted as a moment characterizing the considered random pattern. Knowing one such moment provides a characteristic of the studied object but will not completely describe it. Nevertheless, it provides an important general feature of the entire analysed pattern, and is therefore valuable for data analysis and interpretation \citep{baddeley2006}.
	
	The observation window $\mathit{W}$ serves as a sampling frame in a larger overall study area. This includes an inherent bias for the estimation of $\mathit{F}$ due to the edge effect wherever the borders obscure the actual distance to an element that lies outside $\mathit{W}$. Based on the analogy with the estimation of a survival function, the distance of a reference point $\mathit{u}$ to $\mathit{X}$ is assumed to be right-censored by its distance to the boundary of $\mathit{W}$. Several estimators built on these ideas are available in the literature \citep{Baddeley1999}. Here we have implemented the estimator of $F$ given by Foxall \& Baddeley \citep{Foxall2002}.

\begin{figure} [!htbp]
		\includegraphics[width=0.9\textwidth]{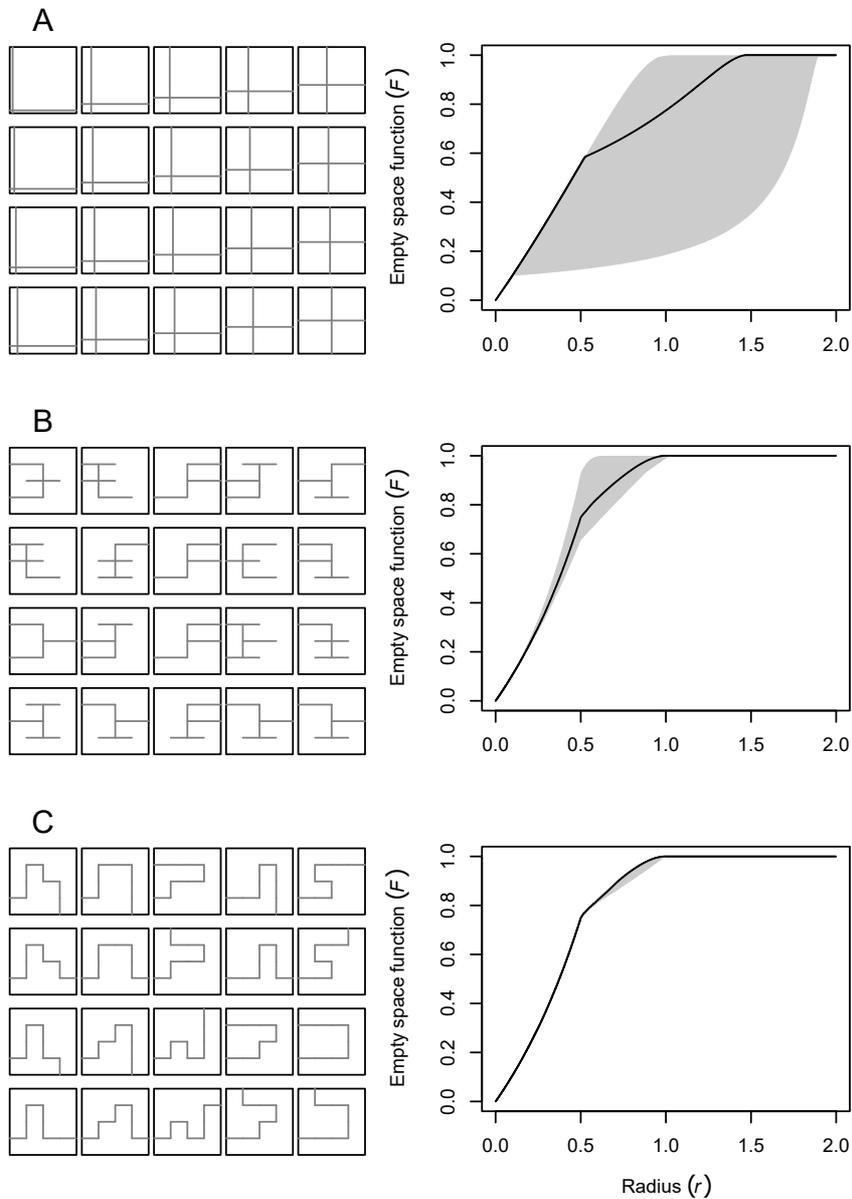}
		\centering
		\caption{\label{fig:patterns} Computation of the empty epace function $F$ for three toy models of line patterns: (A) "main roads", (B) "secondary roads" and (C) "rivers". All patterns have the same line-length of 8 and observation windows of 4$ \times $4; 20 examples for each pattern are shown on the left. The 5-95\% range envelope (grey) and the median (black) are shown on the right. The Y axes are the values of the estimation of $F$, and the X axes the radius $\mathit{r}$. These computations were done using 100 simulations for each model. The R codes for this example are provided in the supplementary information.} 
	\end{figure}

	We applied this type of analysis to our road networks data. The $F$ estimation requires the evaluation of the probability that circles of increasing radii centred on any point in $W$ intersect the line pattern. We used a toy model (Fig.~\ref{fig:patterns}) to demonstrate the use of $F$ for linear features that can occupy a limited available space in different ways. The considered model simulates line patterns of the same length, but with a different topology. The situations depicted in Fig.~\ref{fig:patterns} are a simple and naive replica of the characteristics of three types of network: main roads, secondary roads and rivers.
	
	We performed multiple simulations for each model. The $F$ function was evaluated for each model simulation on a finite set of values of $r$. This allowed a statistical analysis as in Illian et al. \citep{Illian2008}, enabling the construction of range envelopes for $F$, hence characterizing the linear pattern corresponding to each model. In order to synthesise all the information gathered using this statistical approach, we consider the median curve obtained from these envelopes. The results obtained from 100 simulations are given in Fig.~\ref{fig:patterns}. Clearly, the length of the line patterns is not the only element characterizing such an object. The $F$ function appears to successfully integrate more information related to the entire line network topology.
	
	\subsection{Sampling and statistical analyses}
	Roadless space in the study area was calculated inside randomly placed, square observation windows. The size of each window was 30$ \times $30 km, this length being set as twice the maximum distance to a road of any point in the study area. For each year we ran 10~000 replications and kept only those windows that had at least 80\% of their area in the same management/protection status category. This gives for each year and category a number of samples $\mathit{n}$ between 500 and 1000. For each replicated window we derived $\mathit{F}$. The $r$ domain for the application of the empty space function is the interval given by $[0,15]$ km which we divided into discrete steps of 0.2 km. Now, for year and category, for the fixed values of $r$, we have $\mathit{n}$ corresponding values for $F$. Empirical quantiles were computed for each $r$ value and, finally, the median was considered. Therefore, by considering all the medians for all the radius values, we get a median curve $F$ for each management/protection category. This median is denoted $\tilde{F}$.
	
	We next compared these median curves between two management/protection categories $\tilde{F}_{A}$ and $\tilde{F}_{B}$. The obtained functions $\tilde{F}$ are not guaranteed to be empirical distribution functions, hence an alternative to the Kolmogorov-Smirnoff test should be used. We therefore developed an alternative representation of these values: let us consider the set of  differences $d_i = \tilde{F}_A(i) -\tilde{F}_B(i)$. Assuming the sample $d_i, i=1\ldots,n$ to be the realisation of some independent and identically distributed random variables, under the hypothesis that $\tilde{F}_A = \tilde{F}_B$, we have $E[d] = 0$. We tested this statement using a one-sided \textit{t} test. Here we interpret the \textit{t} test statistics as the change intensity in $\tilde{F}_B$ compared to $\tilde{F}_A$. The drawback of this test is that if the null hypothesis $H_0$ is not rejected, this does not mean that $\tilde{F}_A$ and $\tilde{F}_B$ are equal. Therefore, our focus lies on those cases where $H_0$ is rejected. For all $E[d] \neq 0$ we compared $d_i$ between multiple categories using pairwise \textit{t} test with pooled standard deviations and Holm-adjusted \textit{P} values for all possible comparisons \citep{Holm1979}.
	
	All analyses were carried out in R \citep{RCoreTeam2014} using the “spatstat” package \citep{Baddeley2005}. The R codes implementing the present methodology are available by simply sending an email to the first author of this paper.
	
	\section{Results}
\begin{figure}[!htbp]
		\centering
		\includegraphics[width=0.65\textwidth]{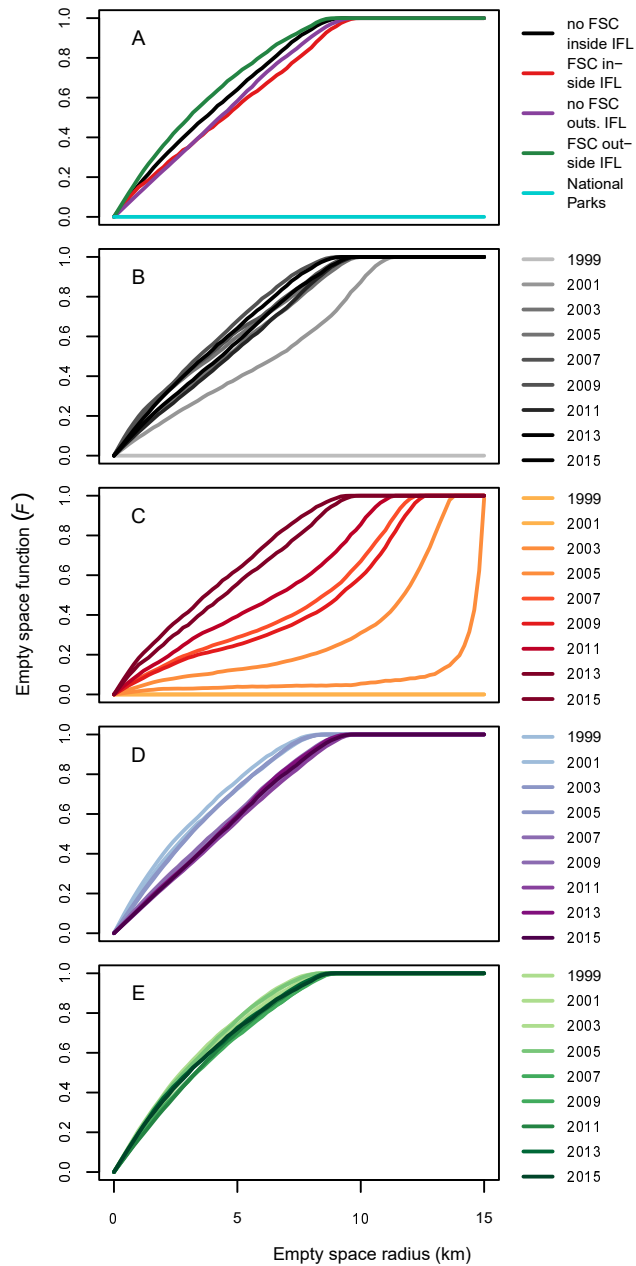}
		\caption{\label{fig:ES} Median curves of the Empty space function \textit{F} (probability that from any point in the observed domain there is a road at distance \textit{r}) against \textit{r}. (A) shows the situation in 2015 for the five categories, (B) non-certified concessions inside intact forest landscapes (IFL), (C) FSC-certified concessions inside IFL, (D) non-certified concessions outside IFL and (E) FSC-certified concessions outside IFL. Panels (B)-(E) show two-year steps from 1999 to 2015 (shading from light to dark with change over time).} 
	\end{figure}

	In 2015 roadless space was very similar inside and outside IFL as well as in certified and non-certified areas outside national parks (Fig.~\ref{fig:ES}~A). However, the median curve of the \textit{F}-function for FSC concessions outside IFL was slightly higher than the others, indicating that this category had less roadless space. For all of the study area outside the protection of national parks we found a maximum distance of 9.8~km from any point to a road. This means that the maximum possible distance between two roads is 19.6~km. No roads were detected inside national parks (Fig.~\ref{fig:ES}~A). Intact forest landscapes were defined based on road networks in the year 2000 \citep{Potapov2008} and accordingly the \textit{F}-function that we calculated also consistently showed values of zero for all concessions inside IFL in 1999  (straight lines in Fig.~\ref{fig:ES}~B and C). Subsequently, there was a clear decrease of roadless space inside IFL over time (Fig.~\ref{fig:ES}~B,C). Specifically, non-certified concessions (Fig.~\ref{fig:ES}~B) showed a rapid decrease from 1999 to 2003 but in subsequent years remained on a similar level to the concessions outside IFL (Fig.~\ref{fig:ES}~D) with a slight increase after 2009. Concessions inside IFL that have been certified since 2006 showed a slower but continuous decrease in roadless space, indicated by regularly increasing curves (Fig.~\ref{fig:ES}~C). By 2013 these areas had reached the same level as all other concessions. Roadless space in certified and non-certified concessions outside IFL changed little between 1999 and 2015 (Fig.~\ref{fig:ES}~D and E). Only for non-certified concessions outside IFL (Fig.~\ref{fig:ES}~D) did the curves show a slight increase in roadless space.
	
	Extracting the probability of encountering the nearest road within 5~km for each year illustrates continuous trends over time. In 2001 the probability of encountering a road within 5~km distance was almost 80\% outside IFL and 30\% or less inside IFL (Fig.~\ref{fig:boxplots}~A). By 2015 it was around 60\% for all management categories except for certified concessions outside IFL which showed a higher probability of around 80\% (Fig.~\ref{fig:boxplots}~B). In general, roadless space inside IFL decreased dramatically, with certified concessions showing a continuous trend until 2015. Outside IFL roadless space remained at a similar level, with a slight but continuous increase for non-certified concessions (Fig.~\ref{fig:boxplots}~C-F).
	
\begin{figure}[!htbp]
		\centering
		\includegraphics[width=0.8\textwidth]{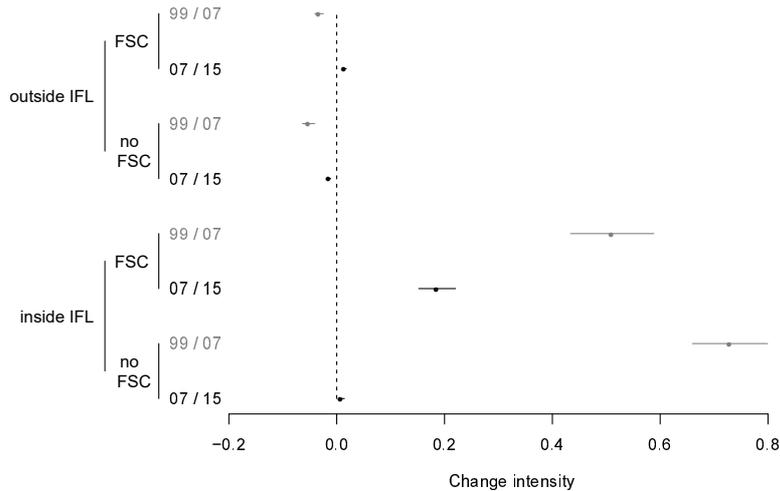}
		\caption{\label{fig:ratio} \textit{t} test statistics and 95\% confidence intervals as measures of change intensity in roadless space (\textit{F}) during two eight-year time periods (1999-2007, grey, and 2007-2015, black), for four management categories: logging concessions with and without Forest Stewardship Council certification (FSC and no FSC), located inside and outside intact forest landscapes (IFL). No overlap of the confidence intervals with zero indicate significant changes between the two years.} 
	\end{figure}

	Comparing changes in roadless space in two longer (eight-year) time steps highlights the contrasts amongst all four management categories. Inside IFL, roadless space decreased by more than half during 1999-2007. Then during 2007-2015 this decrease continued (at a lower rate) in the concessions that were certified from 2006 onwards, whereas it stagnated in non-certified concessions (Fig.~\ref{fig:ratio}). The change in roadless space from 2007 to 2015 was significantly different in certified concessions inside IFL from all other categories, which remained at a similar level (Table~\ref{tab:test}). 
	Intensities of change in roadless space were much less outside IFL. There was a slight increase during 1999-2007, then during 2007-2015 the change intensity was close to zero with only a slight increase in non-certified concessions and a decrease in the concessions that were certified from 2006 onwards (Table~\ref{tab:meanCI}).
	
	\section{Discussion}
	Areas that were classified as intact forest landscapes (IFL) in 2000 are likely to be those which had been subject to least recent logging activity. With the ongoing process of governments leasing concessions for areas of unlogged forest, our result that roadless space greatly decreased inside these IFL during 1999-2007 was expected. However, that this process continued during 2007-2015 in concessions certified by FSC since 2006, whereas it stopped in non-FSC concessions, is surprising. Selective logging in the region focuses on only a few commercial timber species of high market value and expanding logging into previously unlogged areas is more lucrative than repeating logging in the same area \citep{Laurance2000}. However, road construction is expensive and the length of roads built is limited by the capital resources of operating companies. FSC guidelines restrict logging intensity and impose strict requirements to reduce road-related impacts such as poaching, erosion and disturbance of watercourses \citep{FSC2012}, while recommendations about road networks only suggest avoiding “poorly designed patterns of roading” \citep{FSC2010}. This commonly means reducing road lengths to a minimum while still allowing sufficient access to the timber resource \citep{Picard2006, Gullison1993}. However, recent discussions about certified logging in IFL highlight the importance of maximising conserved roadless space as a new component in sustainable forest management. The key consideration in choosing between different road layouts should be to retain undissected blocks of forest and this requires control of the position of permanent roads in the overall forest landscape.
	
	\subsection{Roadless space in certified forests decreased}
	The only net losses of roadless space that we recoded since 2007 took place in FSC-certified concessions. This means that in all other areas, roads that disappeared outweighed those that were newly built. This is linked with the relatively short time that logging roads remain open after abandonment \citep{Kleinschroth2015}. We assume that road building in each concession is related to the capital resources of the operating companies. The intensity of forest exploitation in the region is fluctuating and was strongly affected by the global economic crisis in 2009 \citep{Karsenty2010} when a reduction in the volume of timber exports resulted in temporary shutdowns of logging activities. However, official wood production volume data (available at http://www.observatoire-comifac.net) shows that in subsequent years exploitation increased progressively. This was especially true for the two companies IFO and CIB, which operate four of the largest concessions in the study area, all of which obtained FSC certification between 2006 and 2009. The annual wood production volume increased by 45\% from 2009 to 2013 in the IFO concession and by 24\% in the CIB concessions \citep{DeWasseige2015}. There are several potentially confounding factors that may have influenced this. Well-capitalized companies may have adopted both FSC certification and efficient means of exploitation, leading to a higher density of roads, whereas less capitalized companies did not. Logging history also differs between concession areas. In concessions that have previously been logged, there is less likelihood of yielding high timber volumes in the present cycle and therefore companies may be less likely (a) to accept the costs of FSC certification or (b) to build many new roads. 
	
	The protection of IFL has to be approached at a larger scale than individual logging concessions. Given that the extent of most IFL exceeds the size of individual concessions by far, there is still the potential to retain much greater roadless areas than can be achieved by a single operating company within one concession. The most important factor driving the reduction of roadless space is the position of roads in the overall forest landscape within which this space is located. Roads built closer to the edge have a less fragmenting effect than those in the center. Bigger concessions and those located in the center of the overall forest landscape therefore have a more important role in retaining a large contiguous area that is road-free. Annual felling areas have to be outlined in management plans before exploitation and are the determining factor for where new roads are built (Fig.~\ref{fig:aac})\citep{Cerutti2008}. So far, there is no limitation of the maximum size and position that these areas are allowed to have. Instead, FSC and other forest management guidelines focus on the annual allowable cut, the maximum volume of wood per year and minimum size of trees that are harvested \citep{Cerutti2011}. The environmental advantage of reducing logging intensity is equivocal if instead a greater area needs to be exploited (and made accessible by roads) in order to produce the same volume of timber as before \citep{Healey2000}.
	
	\subsection{Roadless space outside IFL is stabilizing}
	Outside IFL there was a low rate of change in roadless space with a small increase sustained during 1999-2015 in non-certified concessions. This represents the saturation of the forest area with permanent roads some time after the commencement of timber exploitation. It is expensive for companies to maintain and control roads that are permanently accessible and they have a strong interest in reducing the extent of open roads to the minimum that is necessary for their operational efficiency. Overall, more than 80\% of all logging roads are closed and abandoned after a short period of harvesting \citep{Kleinschroth}, a practice in accordance with recommendations for good forest management \citep{Dykstra1996}. Our results indicate that, inside non-certified concessions located in areas with a longer logging history (i.e. outside IFL), in recent years the ratio between abandoned and newly created roads led to a slight increase in roadless space.
	
	\subsection{Minimise the dissection of tropical forests by permanent roads}
	We suggest to extend forest road building recommendations by taking into account the spatial layout of the road network (and thus the conserved roadless space) on regional scale. Based on our findings of limited road persistence \citep{Kleinschroth2015} we suggest that logging roads should be managed as transient elements in the landscape that affect only a small part of the overall area of forest at a given point in time. Road networks should be planned so that the majority of the forest area remains inaccessible to damaging human activity (i.e. as roadless space) at all times, but the exact location of this roadless space may shift over time (Fig. \ref{fig:strategy2}). Other than the theoretical model of conservation concessions, where after a first cut the concession area becomes protected \citep{Gullison2001}, logging companies should be held responsible for conserving the integrity of the forest except for the small portion where logging takes place every year. Key to this integrity will be the planning and management of the road network. A particular danger is posed by permanent access roads that often transect forest areas, thus keeping the core of the forest open to permanent threats, providing pressures exist. Wherever necessary, these roads should be closed effectively and replaced by less fragmenting ones in the periphery of the forest. We suggest this as a principle for all logging road management in tropical forests. 
	
	\subsection{Conclusions}
	The protection of intact forest landscapes (IFL) is currently being incorporated into FSC certification policy. Given that the extent of most IFL exceeds the size of individual concessions and frequently crosses country borders, this process affects multiple stakeholders. Based on our findings we suggest that control of the spatial arrangement of permanent roads is an essential requirement to protect IFL. We recommend that measures to reduce the impacts of selective logging should not only be based on amounts of timber extracted per time and area but also include the size of forest areas that remain undissected by roads. The preservation of large connected areas of roadless space needs to become a top priority in forest management and this can only be achieved by effectively monitoring the spatial arrangement of roads that are permanently accessible for which we advocate the use of the empty space function developed in this paper.

\section*{Acknowledgments}
		This study was funded by the European Union Erasmus Mundus joint doctorate programme FONASO. We thank M. N. M. van Lieshout for essential advice on the methodology and Alain Karsenty for helpful comments on the manuscript.

\Urlmuskip=0mu plus 1mu\relax
\bibliographystyle{plainnat}

\clearpage
	\section*{Supplementary information}
	\renewcommand{\thefigure}{S\arabic{figure}}
	\renewcommand{\thetable}{S\arabic{table}}
	\setcounter{figure}{0}
	
	\begin{table}[!htbp]
		\def\arraystretch{1.5} 
		\small
		\caption{\label{tab:test} Changes in the median empty space function $\tilde{F}$ for the road networks in each of four management categories during two time periods (years 1999-2007 and 2007-2015). \textit{P} values for \textit{t} tests comparing $\tilde{F}$ of different individual categories are shown (\textit{P} values $<$ 0.05 are treated as significant). IFL = intact forest landscapes in 2000; FSC = certified by the Forest Stewardship Council from 2006 onwards. National Parks were not included because no change was detected over time.}
		\begin{tabular}{ llcccc }
			\hline
			Period  & & IFL, FSC &  IFL, no FSC & no IFL, FSC \\ 
			\hline
			1999 - 2007 & IFL, no FSC & $<$ 0.001 &  &  \\ 
			& no IFL, FSC & $<$ 0.001 & $<$ 0.001 &  \\ 
			& no IFL, no FSC & $<$ 0.001 & $<$ 0.001 & 0.597 \\ 
			2007 - 2015 & IFL, no FSC & $<$ 0.001 &  &  \\ 
			&no IFL, FSC & $<$ 0.001 & 0.637 &  \\ 
			&no IFL, no FSC & $<$ 0.001 & 0.138 & 0.067 \\ 
			\hline
		\end{tabular}
		
	\end{table}
	
	\begin{table}[!htbp]
		\def\arraystretch{1.5} 
		\caption{\label{tab:meanCI} \textit{t} test statistics and 95\% confidence intervals comparing changes in roadless space over time between management categories. Inputs are the median values of the empty space function $\mathit{F}$ for the road networks in 1999, 2007 and 2015. The \textit{t} statistic is interpreted as the change intensity. IFL = inside intact forest landscapes in 2000; FSC = certified by the Forest Stewardship Council from 2006 onwards; no FSC = not certified by FSC; CI = confidence intervals. National Parks were not included because no change was detected over time.}
		\begin{tabular}{llcccc}
			\hline
			Category &  & Compared years & Change intensity & 2.5\% CI & 97.5\% CI \\ 
			\hline
			outside IFL & FSC 	& 1999 / 2007 & -0.033 & -0.040 & -0.025 \\ 
			&  			& 2007 / 2015 & 0.014 & 0.011 & 0.018 \\	 
			& no FSC 		& 1999 / 2007 & -0.052 & -0.064 & -0.041 \\ 
			&  			& 2007 / 2015 & -0.014 & -0.017 & -0.011 \\		 
			inside IFL & FSC 	& 1999 / 2007 & 0.511 & 0.434 & 0.588 \\ 
			&  			& 2007 / 2015 & 0.186 & 0.152 & 0.220 \\		 
			& no FSC 		& 1999 / 2007 & 0.730 & 0.660 & 0.799 \\	 
			&  			& 2007 / 2015 & 0.009 & 0.003 & 0.014 \\		 
			\hline
		\end{tabular}
	\end{table}
\clearpage
\begin{figure}[!htbp]
	\includegraphics[width=0.8\textwidth]{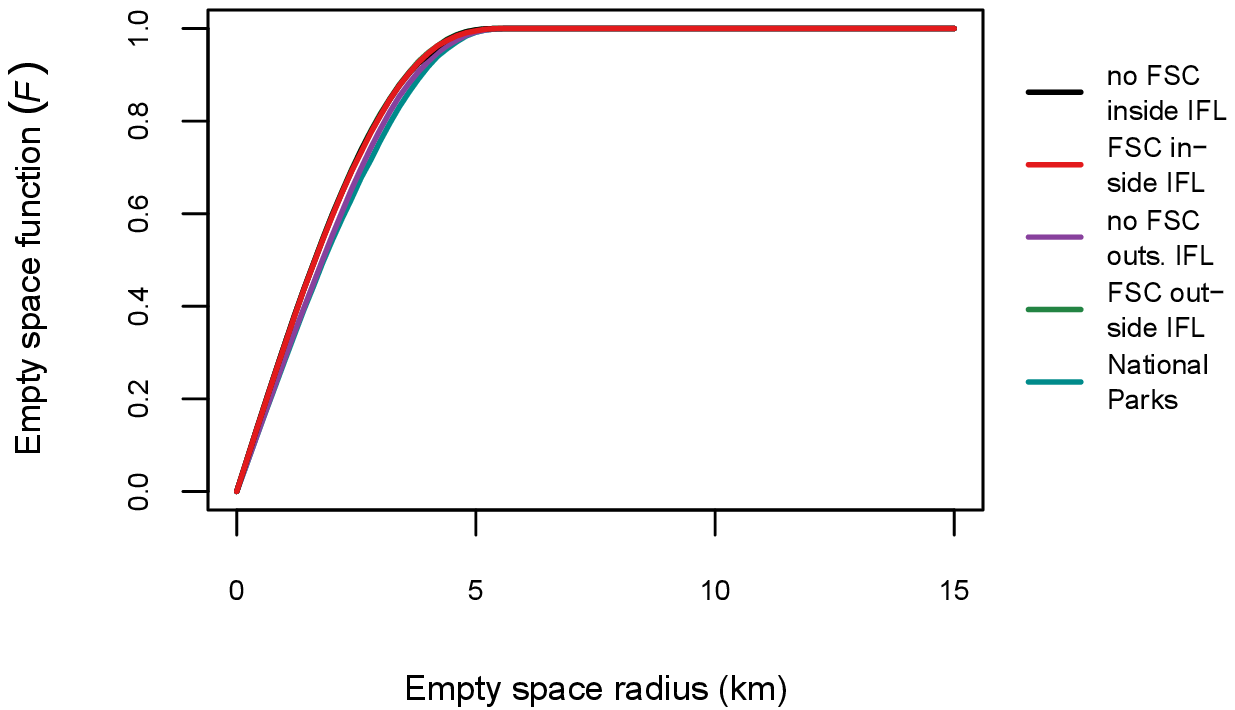}
	\caption{\label{fig:riv} River network analysis: median curves of the empty space function $F$ (probability that from any point in the observed domain there is a river at distance $r$) against $r$ in the five different management/protection categories.}
\end{figure}
	
	\begin{figure}
		\includegraphics[width=1\textwidth]{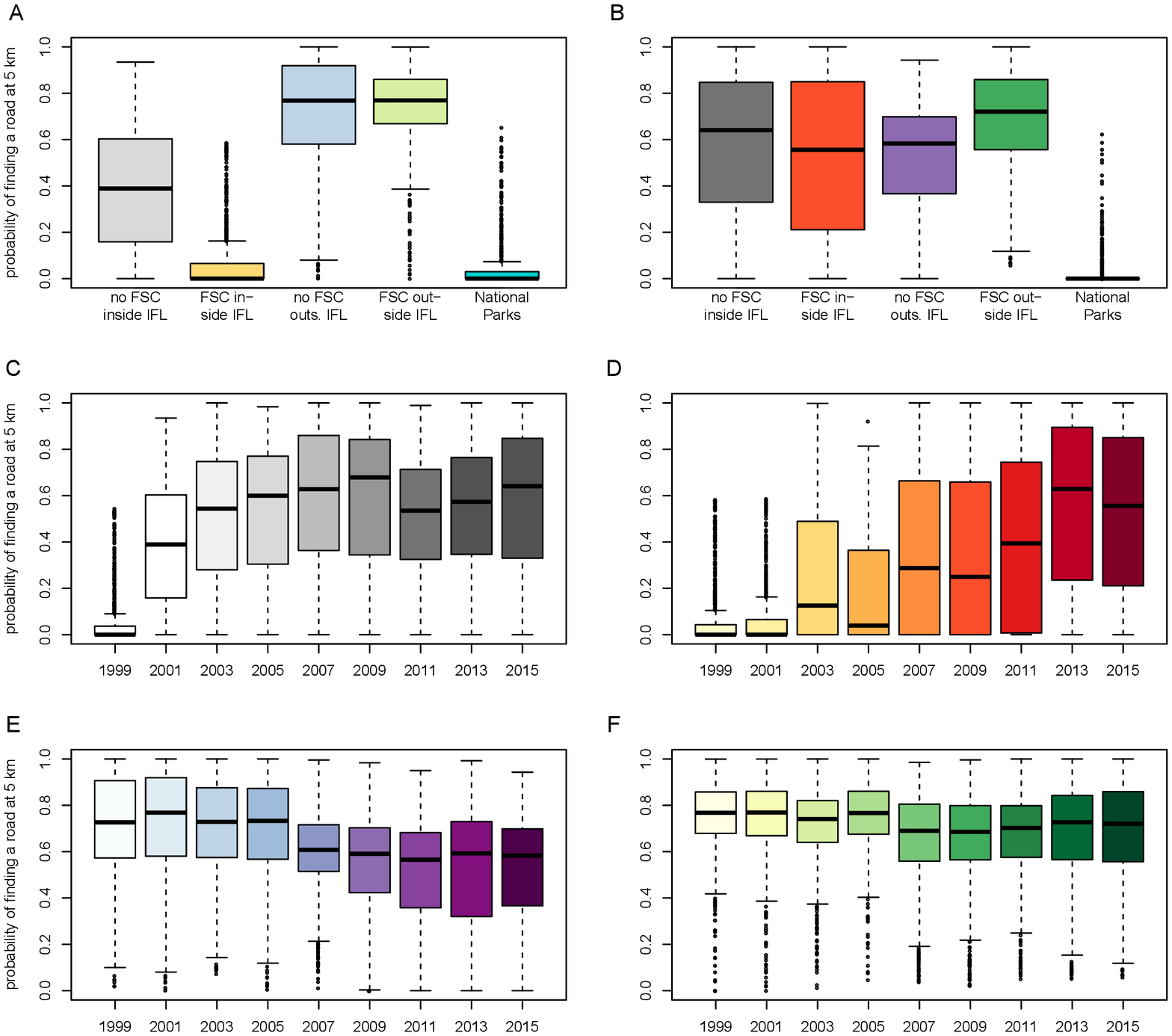}
		\caption{\label{fig:boxplots} Boxplots for the envelopes of Empty-Space curves for the probability of intersecting a road within a distance of 5~km. (A) is the combined situation in 2001 and (B) in 2015 for five different forest management/protection categories. (C) Non-certified concessions inside intact forest landscapes (IFL), (D) FSC-certified concessions inside IFL, (E) non-certified concessions outside IFL and (F) FSC-certified concessions outside IFL. Panels (C)-(F) show two-year steps from 1999 to 2015.} 
	\end{figure}
	
	\begin{figure}[!htbp]
		\centering
		\includegraphics[width=1\textwidth]{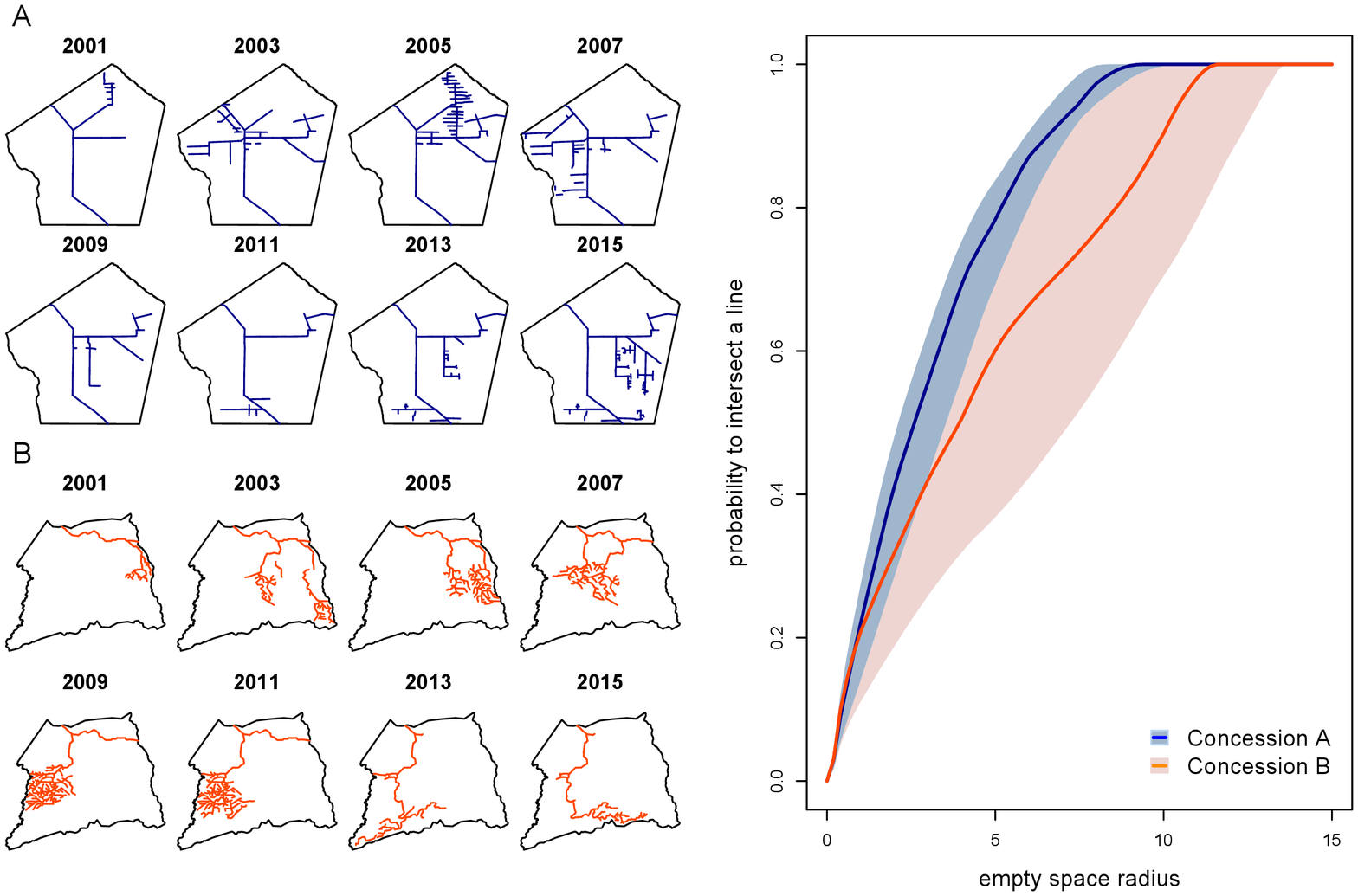}
		\caption[Effects of road building strategies in two concessions]{\small \label{fig:strategy2} The effect on roadless space at two-year intervals during the period 2001-2015 of two contrasting road network development strategies that were implemented during timber harvesting, respectively, in two logging concessions one (A) in Republic of Congo and one (B) in Cameroon. The two concessions were selected to be matched as closely as possible in logging history, site characteristics and area, however their boundaries are slightly modified for this modelling so that they have the same area (1227 km2) and regular shape with a diagonal extent of ca. 50~km. The average road length for the eight sample years is 139 km for concession A and 144~km for concession B. The curves on the right show the median Empty-Space Function for all years as a solid line and the 5-95\% ranges amongst the eight sample years are shaded, indicating that overall roadless space was greater in concession B.} 
	\end{figure}

\end{document}